\documentclass[9pt,twocolumn,twoside]{osajnl}
\usepackage{tabularx}
\usepackage{booktabs}
\usepackage{multirow}
\usepackage{rotating}

\journal{jocn} 

\setboolean{shortarticle}{false}

\title{A Directly Modulated Laser Platform for High-Dimensional Quantum Key Distribution}

\author[1,2,3,*]{Yang Zhou}
\author[1,2,3,*]{Xing-Yu Zhou}
\author[1,2,3]{Shu-Fan Wu}
\author[4]{Qiang Zeng}
\author[4]{Zhi-Liang Yuan}
\author[1,2,3,$\ddagger$]{Qin Wang}

\affil[1]{Institute of Quantum Information and Technology, Nanjing University of Posts and Telecommunications, Nanjing 210003, China}
\affil[2]{Key Lab of Broadband Wireless Communication and Sensor Network Technology, Ministry of Education, Nanjing University of Posts and Telecommunications, Nanjing 210003, China}
\affil[3]{Telecommunication and Networks, National Engineering Research Center, Nanjing University of Posts and Telecommunications, Nanjing 210003, China}
\affil[4]{Beijing Academy of Quantum Information Sciences, Beijing 100193, China}

\affil[*]{These authors contributed equally to this work.}
\affil[$\ddagger$]{qinw@njupt.edu.cn}

\begin{abstract}
High-dimensional quantum key distribution (HD-QKD) offers a promising approach to enhance secret key rates beyond conventional binary-encoded QKD, addressing the growing demand for secure data transmission. However, the practical application of most HD-QKD systems has been hindered by their complexity, as they require the preparation and detection of quantum states in large Hilbert spaces. Here, we design and experimentally realize a directly modulated laser platform for HD-QKD. It operates at a repetition rate of 312.5 MHz, yielding a remarkably simple and scalable architecture. Through which, we achieve a record transmission distance of 250 km for HD-QKD, demonstrating its feasibility for long-distance quantum communication. Furthermore, we witness that the four-dimensional states outperform their two-dimensional counterpart in secret key rate, highlighting the practical advantage of high-dimensional encoding. This simple and scalable approach shows strong potential for chip-scale integration.
\end{abstract}

\setboolean{displaycopyright}{false} 

\begin{document}

\maketitle

\section{Introduction}
Quantum key distribution (QKD) is one of the most mature and practical applications of quantum communication, enabling two distant users to share information-theoretically secure keys through the transmission of quantum states over optical channels \cite{QKD1,QKD2,QKD3}. Unlike classical cryptography \cite{trodiction1,trodiction2}, its security stems from the fundamental principles of quantum mechanics \cite{ANQUAN1,ANQUAN2} rather than assumptions about computational hardness. Over the past decades, QKD has evolved from proof-of-principle demonstrations to field deployments, and is widely regarded as a cornerstone for building future quantum networks and a global-scale quantum internet. However, despite remarkable progress, practical QKD systems still face two critical challenges: the limited secret key generation rate and the strong sensitivity to channel noise \cite{QUESTION}. The limitations directly impact both the efficiency and robustness of QKD in real-world environments, particularly over long-distance fiber links or in metropolitan networks.

These challenges are closely related to the inherent limitations of conventional two-dimensional (2D) encoding protocols. In such systems, each photon encodes at most one bit of information, fundamentally constraining the achievable key rate under channel losses. Furthermore, the noise tolerance of 2D schemes is intrinsically limited \cite{ANQUAN2,NOISE}. For instance, in the BB84 protocol under coherent attacks with one-way reconciliation, the maximum tolerable quantum bit error rate (QBER) is about 11.0\% \cite{2D11}. Once the QBER exceeds this threshold, no secure key can be distilled. In practical systems, noise can arise from diverse sources, including stray light, various scatterings in channels, or even from eavesdropping attacks, making the low tolerance of 2D systems a severe bottleneck \cite{noise1,noise2,noise3}. To overcome these issues, high-dimensional (HD) QKD has been proposed and extensively analyzed, where quantum information is encoded in qudits belonging to a larger Hilbert space \cite{HD1,HD2,HD3}. By increasing the dimension $d$, each photon can in principle carry $\log_2 d$ bits of information, thereby boosting secret key rates. At the same time, HD-QKD offers significantly higher resilience to noise: for example, the QBER threshold can be raised to about 18.9\% and 24.7\% in four-dimensional and eight-dimensional protocols, respectively \cite{4D18,HD1}. These features make HD-QKD a promising route for practical, high-capacity and noise-robust quantum communication.

To realize HD-QKD, the main physical degrees of freedom (DoF) that have been explored include orbital angular momentum (OAM) \cite{OAM1,OAM11,OAM12,OAM13,OAM2}, path encoding \cite{path1,path2,path3,path4,path5}, time-bin encoding \cite{timebin1,timebin2,timebin3,HDCOW}, and time-energy encoding \cite{timeengry1,timeengry2,timeengry3,timeengry4,timeengry5,timeengry6}. In addition to these primary DoFs, several hybrid encoding schemes have also been investigated, such as OAM–polarization \cite{OAMpol}, OAM–spin angular momentum (SAM) \cite{OAMSAM}, time-bin–path \cite{timepath} and time-bin and polarization hyperentanglement \cite{timebinpol}. Among them, time-bin encoding of weak coherent laser pulses is particularly attractive for fiber-based quantum communication due to its intrinsic compatibility with standard telecom infrastructure \cite{HDCOW}. Demonstrations of HD time-bin encoding have shown substantial improvements in key rates and noise tolerance, including a record secure key rate of 26.2 Mbit/s using a four-dimensional (4D) protocol resilient against coherent attacks \cite{timebin1}. However, these approaches often involve considerable complexity at the transmitter side: as the dimensionality increases, quantum-state preparation becomes more demanding in terms of modulation precision, phase stability, and temporal synchronization. Generating mutually unbiased high-dimensional states typically requires multiple modulators and carefully calibrated optical paths to ensure high-fidelity encoding. On the receiver side, high-dimensional measurements also demand intricate interferometric setups for full state projections. Overall, both the state preparation and measurement processes become increasingly challenging as the system dimension grows, posing significant obstacles to practical and scalable HD-QKD implementations.

Here, we address these challenges by developing a directly modulated laser platform for high-dimensional BB84 QKD using injection-locked sources with time-bin and phase encoding. This design eliminates the need for external intensity and phase modulators, requiring only minimal optical components and significantly reducing transmitter complexity and stabilization overhead. Using this platform, we experimentally achieve 4D QKD over 250~km of fiber—the longest distance demonstrated for high-dimensional time-bin schemes to date. Furthermore, we observe that four-dimensional encoding delivers a higher secret key rate than its two-dimensional counterpart under identical hardware conditions, demonstrating the practical advantage of high-dimensional encoding. Our scheme also enables existing 2D systems to be readily upgraded to high-dimensional operation with only a minimal extension on the receiver side. This simplified architecture not only facilitates practical and scalable deployment but also shows strong potential for chip-scale integration.

\section{HIGH-DIMENSIONAL CODING PROTOCOL}
In discrete-variable quantum key distribution (DV-QKD), information is encoded into discrete degrees of freedom of single photons. To implement the protocol, two mutually unbiased bases (MUBs) are defined and randomly selected by the sender (Alice), while the receiver (Bob) independently chooses one of the two bases for measurement. After transmission over the quantum channel, a basis reconciliation step is performed, followed by error correction and privacy amplification to distill the final secret key. Within this framework, the two bases play distinct roles: the quantum states of the $Z$ basis are adopted for encoding the raw key bits, while the $X$ basis is employed exclusively for security verification, enabling the detection of eavesdropping through the estimation of error rates.  

A straightforward generalization of the BB84 protocol to 4D time-bin encoding was introduced in Ref.~\cite{timebin1}. In this scheme, the $Z$ basis states correspond to single photons localized in one of four distinct time bins, whereas the $X$ basis states are equal-weight superpositions across all bins, distinguished by different relative phases. Projection onto the $Z$ basis can be performed by a direct time-of-arrival measurement, in analogy with the conventional two-dimensional protocol. However, implementing full projections onto the $X$ basis is considerably more demanding, as it requires a cascade of three interferometers with appropriate time delays together with four single-photon detectors~\cite{timebin1,timefreq}. Alternative approaches \cite{other} rely on even more complex solutions, which further increase experimental overhead and complicate practical implementation. These requirements impose stringent stabilization demands and limit scalability, thereby posing challenges for practical deployment.  

\begin{figure}[ht]
	\centering
	\includegraphics[width=1\linewidth]{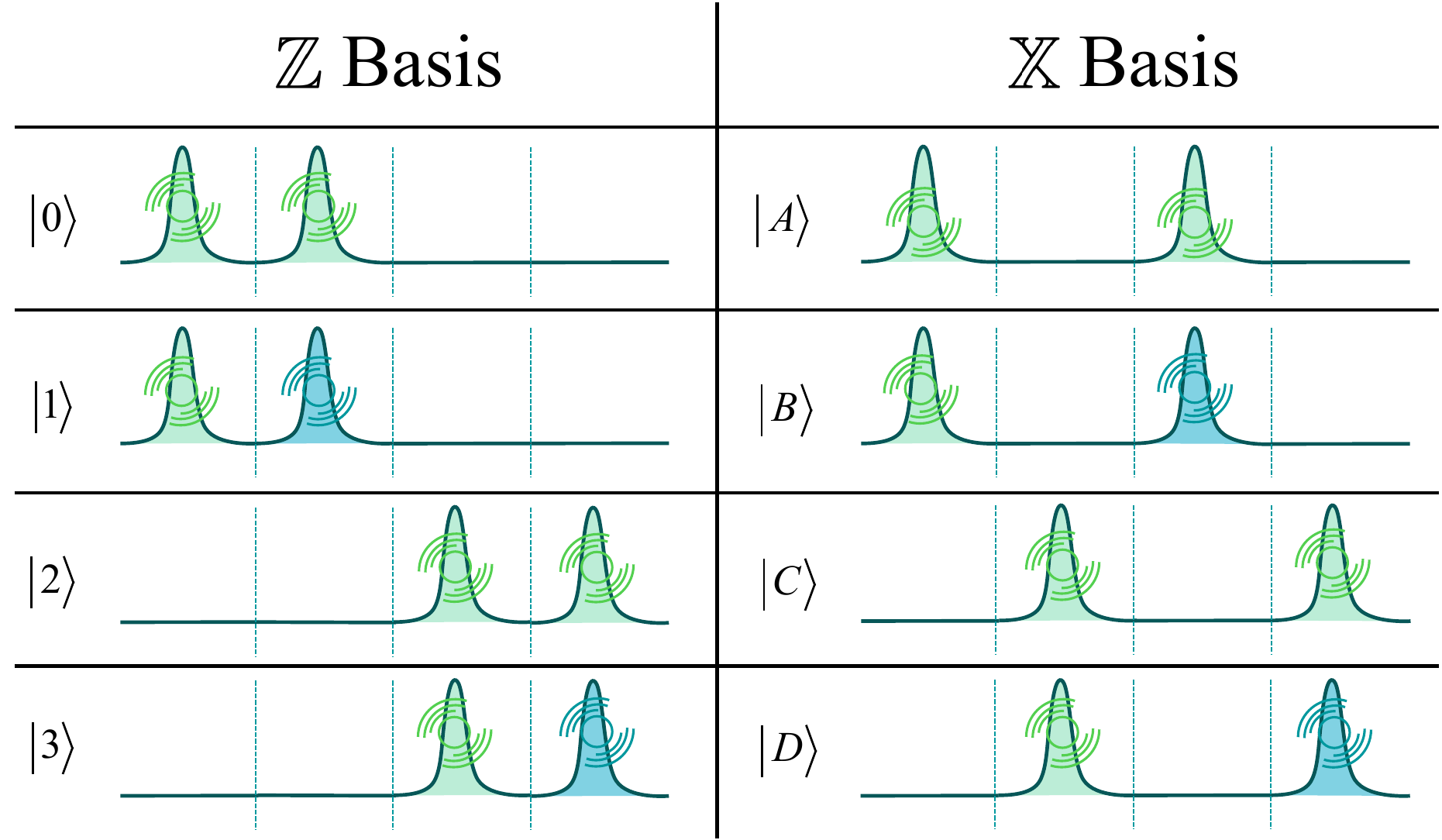}
	\caption{Quantum states involved in the four-dimensional QKD protocol. In the $Z$ basis, quantum states are generated by two pulses in adjacent time slots, whereas in the $X$ basis, quantum states are formed by two pulses separated by one time slot. Within each quantum state, two pulses of the same color indicate a phase difference of 0, while pulses of different colors correspond to a phase difference of $\pi$.}
	\label{fig1}
\end{figure}

To overcome these limitations, we adopt the simplified high-dimensional encoding scheme reported in Ref.~\cite{timebin3}. In this approach, the eight generated states, encoding two bits of quantum information, form two mutually unbiased bases, as shown in Fig.~\ref{fig1}.

In the $Z$ basis, each quantum state is constructed from two consecutive time bins. Explicitly, the four orthogonal states are defined as
\begin{equation}
	\begin{aligned}
		|0\rangle &= \tfrac{1}{\sqrt{2}}(|t_1\rangle + |t_2\rangle), \quad
		|1\rangle = \tfrac{1}{\sqrt{2}}(|t_1\rangle - |t_2\rangle), \\
		|2\rangle &= \tfrac{1}{\sqrt{2}}(|t_3\rangle + |t_4\rangle), \quad
		|3\rangle = \tfrac{1}{\sqrt{2}}(|t_3\rangle - |t_4\rangle).
	\end{aligned}
\end{equation}
Here, the information is jointly encoded in the temporal separation and the relative phase between the two pulses. Pulses of the same color correspond to a phase difference of $0$, while those of different colors correspond to a phase difference of $\pi$. 

In the $X$ basis, the construction is similar, except that the two pulses forming a state are separated by one time slot instead of being consecutive. The corresponding four orthogonal states are
\begin{equation}
	\begin{aligned}
		|A\rangle &= \tfrac{1}{\sqrt{2}}(|t_1\rangle + |t_3\rangle), \quad
		|B\rangle = \tfrac{1}{\sqrt{2}}(|t_1\rangle - |t_3\rangle), \\
		|C\rangle &= \tfrac{1}{\sqrt{2}}(|t_2\rangle + |t_4\rangle), \quad
		|D\rangle = \tfrac{1}{\sqrt{2}}(|t_2\rangle - |t_4\rangle).
	\end{aligned}
\end{equation}
By construction, states within each basis are orthogonal, while the two sets form mutually unbiased bases since $|\langle z_n|x_m\rangle|^2 = 1/d$ holds for all $n,m=0,\ldots,d-1$ with $d=4$~\cite{timebin3}.
To mitigate photon-number-splitting (PNS) attacks, we implemented a one-decoy-state method \cite{DECOY1,DECOY2,DECOY3}, in which signal and decoy pulses with mean photon numbers $\mu$ and $\nu$ were randomly modulated. This approach allows accurate estimation of single-photon contributions and ensures the unconditional security of the HD-QKD system under general attacks.

This encoding strategy offers clear advantages in terms of detection. Projection onto the $Z$ basis can be realized with a single $\tau$-delayed interferometer, while projection onto the $X$ basis requires only one $2\tau$-delayed interferometer, where $\tau$ denotes the time-bin duration. Compared with conventional four-dimensional time-bin protocols that typically require cascaded interferometers and multiple detectors, this scheme reduces hardware overhead and stabilization requirements, thereby enhancing its practicality for scalable HD-QKD implementations.  

Nevertheless, realizing such an encoding scheme at the transmitter side in conventional systems is still relatively demanding. Typically, an intensity modulator (IM) is used to carve the optical pulses and a phase modulator (PM) is required to set the relative phases. This configuration increases system complexity, necessitates additional electronic control signals, and can introduce extra insertion losses. To address these limitations, in our implementation we employ a injection-locked technique with directly modulated lasers \cite{SEED1,SEED2}, which greatly simplifies the transmitter design while ensuring high-fidelity state preparation. The details of this approach will be described in the following section.  

On this basis, the secret key length is estimated in a finite-block-size regime with security parameter $\varepsilon_{\mathrm{sec}}$, correctness parameter $\varepsilon_{\mathrm{cor}}$, and block size $N$ in 4D with
On this basis, the secret key length is estimated in a finite-block-size regime using a block size $N$ in the 4D with \cite{timebin3}
\begin{equation}
	\begin{split}
		\ell_{4D} \leq &\, 2 D_{0}^{Z,L} + D_{1}^{Z,L}\left[2-H\left(\phi_{Z}\right)\right] - \lambda_{EC} \\
		&\, - 6 \log_2\left(19 / \epsilon_{\mathrm{sec}}\right) - \log_2\left(2 / \epsilon_{\mathrm{cor}}\right),
	\end{split}
\end{equation}
where  $D_{0}^{Z,L}$  and  $D_{1}^{Z,L}$  are the lower bounds of vacuum and single-photon events in the $Z$ basis, $H(x):=-x \log _{2}(x / 3)-(1-x) \log _{2}(1-x)$ is the Shannon entropy in a four-dimensional Hilbert space, $\phi_{Z}$ is the phase error rate upper bound, $\lambda_{EC}$ is the number of bits that are publicly disclosed during error correction, and $\varepsilon_{\mathrm{sec}}$ and $\varepsilon_{\mathrm{cor}}$ are the secrecy and correctness parameters, set to $10^{-9}$ and $10^{-10}$, respectively \cite{security}.

\section{EXPERIMENTAL IMPLEMENTATION}
\begin{figure*}[ht]
	\centering
	\includegraphics[width=1.0\linewidth]{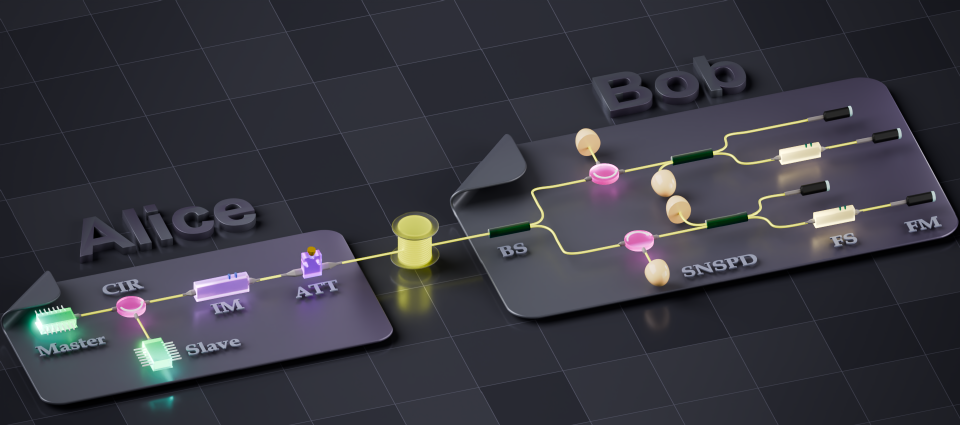}
	\caption{Schematic of our experimental setup.
		Alice employs a directly phase-modulated light source composed of a pair of distributed-feedback (DFB) lasers in a master–slave configuration. Both lasers are driven by the arbitrary waveform generator (AWG), which provides the RF signals required for direct phase modulation. At Bob’s measurement station, two unbalanced Faraday–Michelson interferometers (FMIs) with path-length differences of 800 ps and 1.6 ns, respectively, are used to demodulate the $Z$ basis and $X$ basis pulses. Superconducting nanowire single-photon detectors (SNSPDs) record the interference outcomes. Circulator (CIR); Intensity modulator (IM); Attenuator (ATT); Beam splitter (BS); Fiber stretcher (FS); Faraday mirror (FM).}
	\label{fig2}
\end{figure*}

In our experiment, the most critical component is the modulation of the quantum states. Here, we generate encoded quantum states using directly modulated injection-locked gain-switched lasers \cite{SEED2}. The experimental setup is shown in Fig.~\ref{fig2}. Specifically, Alice employs two independent distributed-feedback (DFB) lasers (Allwave Lasers Photonic Devices Inc.), operating at 1550.12~nm, referred to as the master laser and the slave laser. Both lasers are equipped with thermoelectric coolers (TECs) governed by PID controllers for temperature stabilization, achieving a stability of 0.01~\textcelsius. The arbitrary waveform generator (AWG, Keysight 8190A) produces RF signals. These signals are first amplified by an RF amplifier and then used to drive the lasers with a DC bias. 

\begin{figure*}[ht]
	\centering
	\includegraphics[width=1.0\linewidth]{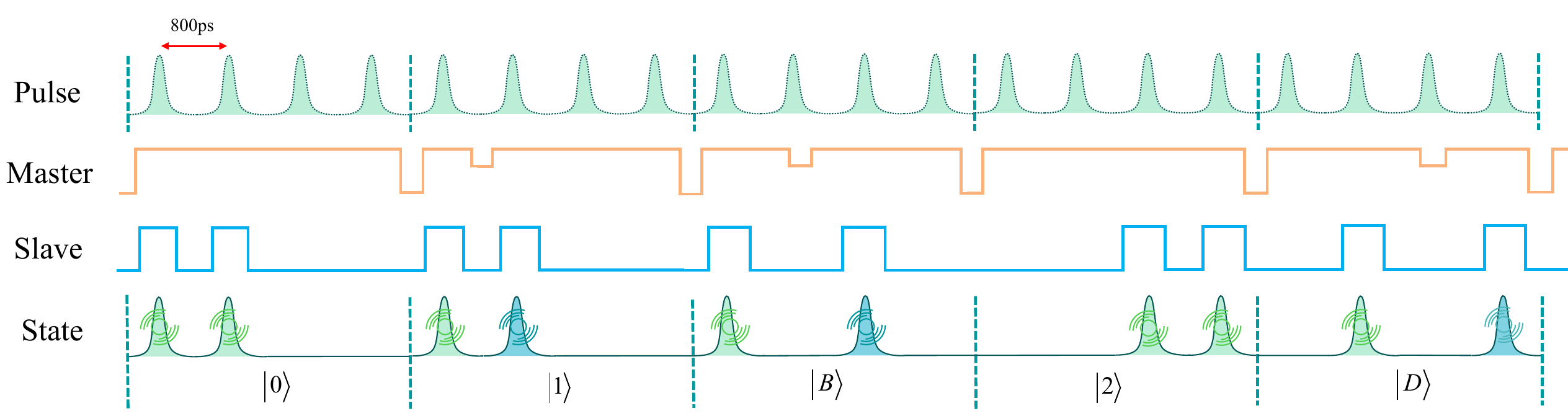}
	\caption{Generation of four-dimensional BB84 states. 
		The top trace shows the optical pulse train with an 800~ps spacing.
		The middle traces depict the electrical drive signals for the slave and master DFB lasers, respectively. 
		The slave laser defines the occupied time bins, while the master laser provides phase modulation through 200~ps amplitude perturbations on its RF drive.
		States of the $Z$ basis are formed by selecting the first–second or third–fourth time bins and applying the appropriate phase shift.
		States of the $X$ basis are similarly produced using the first–third or second–fourth time bins.
		In every state period the global phase is randomized by a freshly seeded master pulse.
		Examples of the generated optical pulses and their corresponding quantum states are shown in the lower panel of the figure.}
	\label{fig3}
\end{figure*}

For 4D-BB84 encoding, the master laser is gain-switched at a clock rate of 312.5~MHz, with an on-time of 3~ns, periodically bringing the laser above threshold and switching off between the pluses so that each master pulse acquires a random phase. The slave laser is gain-switched at specified time bins, with an on-time of 400 ps, as illustrated in Fig.~\ref{fig3}. The RF signals of the master and slave lasers are aligned in time with picosecond precision, ensuring that the slave pulses are correctly seeded by injection from the master pulses. The two lasers are coupled via a polarization-maintaining circulator (CIR). Ultimately, the master pulses, at suitable injection power, seed the slave laser, forming four time bins per state period. Because the slave pulses are seeded by the injected master pulses, they inherit the phase of the injected light, resulting in significantly reduced output-pulse jitter and chirp \cite{SEED1,SEED3}. The relative phases of the slave pulses can be precisely controlled by introducing small amplitude perturbations in the RF drive signal of the master laser. Such a perturbation changes the carrier density in the master laser cavity, which in turn alters the refractive index and causes a transient optical frequency shift \cite{SEED1,SEED4}. Thus, photons generated after this modulation experience a corresponding phase shift. By placing the perturbation in the master laser’s RF signal at a time corresponding to the interval between two slave pulses within a state period, the induced phase shift is imparted to the subsequent pulse, creating a precise relative phase difference between the two pulses.

Amplitude perturbations with a fixed temporal width of 200 ps are applied to the master-laser RF signal to realize the required phase modulation. The perturbation amplitude is adjusted according to the desired phase shift. For the $Z$ basis, the perturbation is applied at the midpoint between the first and second time bins ($t_1$–$t_2$) to encode $\lvert 1\rangle$, while it is omitted for $\lvert 0\rangle$. Similarly, for $\lvert 2\rangle$ and $\lvert 3\rangle$, the perturbation is inserted between the third and fourth time bins ($t_3$–$t_4$), following the same rule. For the $X$ basis, the perturbation is applied midway between $t_1$ and $t_3$ for $\lvert B\rangle$, and between $t_2$ and $t_4$ for $\lvert D\rangle$; in both cases, the absence of perturbation corresponds to $\lvert A\rangle$ and $\lvert C\rangle$, respectively. The global phase of each state is random, as each state period is seeded by an independent master pulse. 

An intensity modular (IM) is employed  to generate two different intensities for the decoy state method. The optical pulses then pass through a 50 GHz optical filter, which suppresses noise and enhances phase coherence (not shown in the setup diagram). Afterward, they are attenuated to the single-photon level using an attenuator (ATT) and transmitted to Bob through low-loss fibers with an average attenuation of less than 0.17 dB/km.

\begin{figure*}
	\centering
	\includegraphics[width=1.0\linewidth]{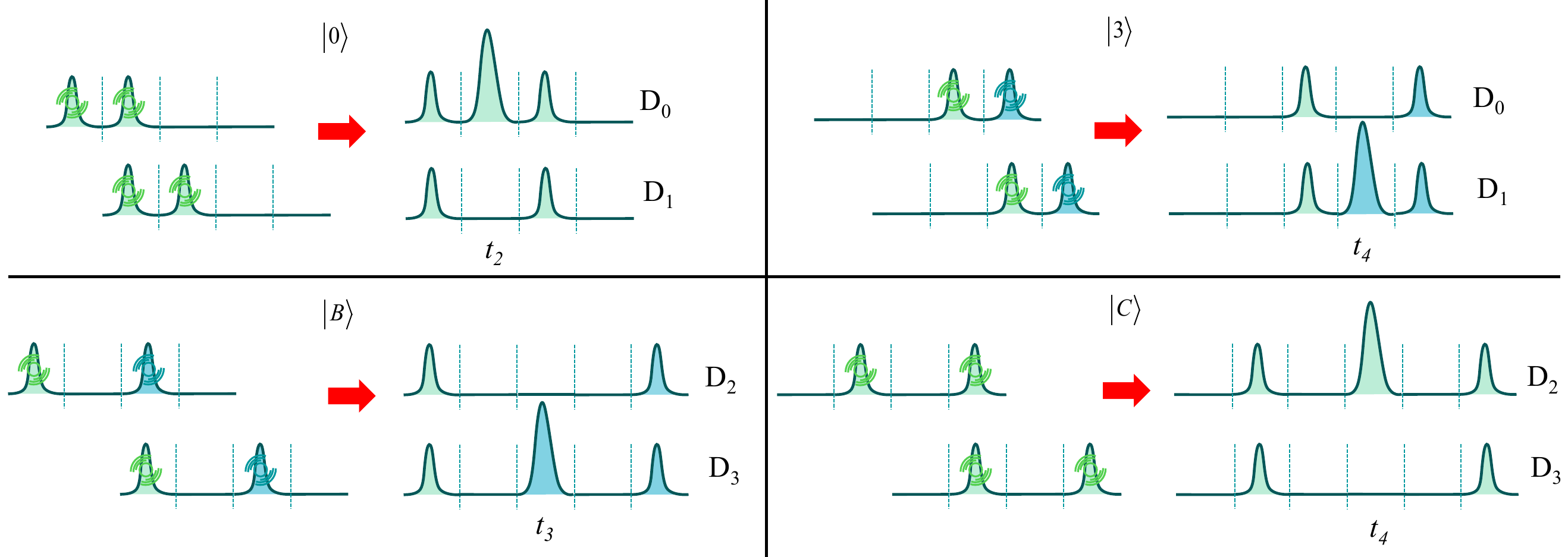}
	\caption{Interference of the four-dimensional quantum states in our scheme. The figure presents representative interference results for four example states, corresponding to the matched-basis measurements, illustrating the interference behavior observed in the relevant time bins. Here, $D_0$ and $D_1$ denote the SNSPDs connected to the 800 ps FMI, and $D_2$ and $D_3$ denote the SNSPDs connected to the 1.6 ns FMI.
	}
	\label{fig5}
\end{figure*}

On Bob’s side, a 50:50 fiber beam splitter (BS) is used for passive bases selection. Decoding of the $Z$ and $X$ bases is realized with two unbalanced fiber Faraday--Michelson interferometers (FMIs) having path-length differences of 800~ps and 1.6~ns, respectively, corresponding to the temporal separations of the pulses in each basis. The interference of the four-dimensional states is observed in the $t_2$ and $t_4$ time bins for the $Z$ basis measurements and in the $t_3$ and $t_4$ time bins for the $X$ basis measurements \cite{timebin3}, as illustrated in Fig. \ref{fig5}. The receiver uniquely determines the projection outcome by observing which time bin clicks and on which detector the click occurs. Fiber stretchers (FS) are mounted on the long arms of both interferometers to provide active phase compensation between Alice and Bob. At the receiver, superconducting nanowire single-photon detectors (SNSPDs) with 75\% detection efficiency, a dark-count rate of 5~Hz.

\section{Result}
Based on the above experimental setup and protocol, we performed the 4D time–bin and phase encoding QKD experiments over fiber links of 200~km and 250~km, respectively. A one-decoy-state scheme was implemented, in which the signal and decoy pulses with mean photon numbers $\mu$ and $\nu$ were used. All experimental parameters, including $\mu$, $\nu$, the emission probabilities $P_{\mu}$ and $P_{\nu}$, and the basis selection probabilities $P_{X}^A$ and $P_{Z}^A$, were numerically optimized to maximize the secure key rate. Since Bob employed a passive basis-choice scheme, the probabilities for selecting the $X$ and $Z$ bases were inherently fixed to $P_{X}^B=P_{Z}^B=50\%$. 

\begin{figure}[t]
	\centering
	\includegraphics[width=1\linewidth]{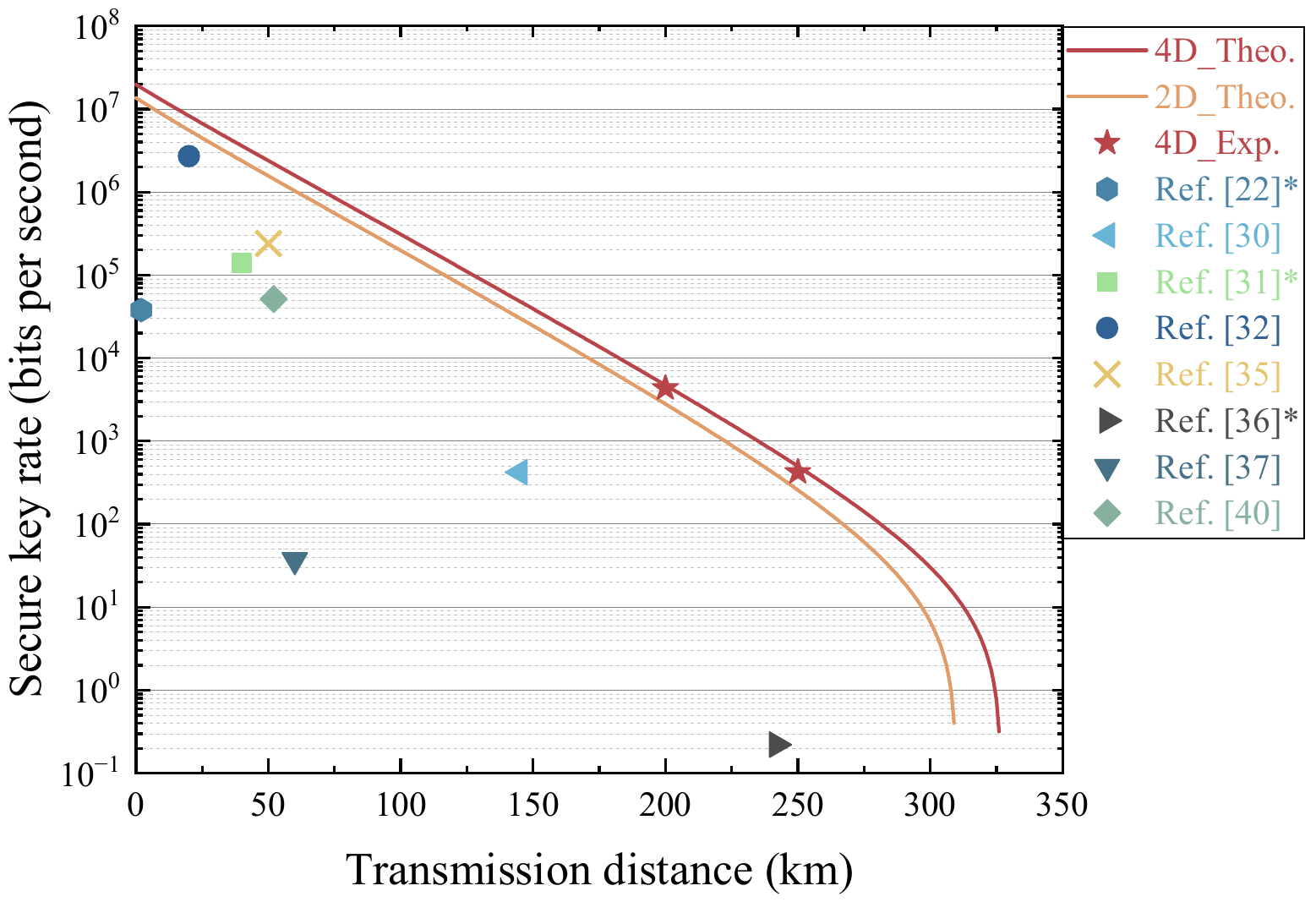}
	\caption{The secret key generation rate is shown as a function of transmission distance, calculated using the parameters of the laboratory experiment, where the repetition rates for the 2D and 4D systems are 625 MHz and 312.5 MHz, respectively. The stars mark the experimentally obtained key rates at distances of 200 km and 250 km, while the other symbols represent the results reported in previous typical high-dimensional experiments. Entries marked with * correspond to asymptotic secret key rates, whereas the other data correspond to finite-key results.}
	\label{fig4}
\end{figure}

Fig. \ref{fig4} presents both the expected and experimental secret key rates of the 4D protocol as a function of transmission distance. The key rate is calculated using the experimental parameters summarized in Table~\ref{tab1}, where $QBER_{Z}$ and $QBER_{X}$ represent the quantum bit-error rates measured in the $Z$ and $X$ bases, respectively, and $\phi_{Z}$ and $\phi_{X}$ denote the corresponding phase error rates in these two bases for signal intensities with mean photon numbers $\mu$ and $\nu$. The measured $QBER_Z$ and $QBER_X$ values are relatively low, remaining below 3.5\%, which confirms the high quality of the quantum state preparation. The main contributions to these error levels arise from the finite extinction ratio of the time-bin pulses and the interference visibility.

\begin{table}[t]
	\centering
	\caption{\label{tab1} Experimental parameters in this work.}
	\label{tab:all-params}
		\begin{tabular}{ccc}
			\hline
			Parameter & 200~km & 250~km  \\
			\hline
			$\mu$ & 0.50 & 0.46  \\
			$v$   & 0.18 & 0.18  \\
			$P_\mu$ & 0.78 & 0.62  \\
			$P_v$ & 0.22 & 0.38  \\
			\hline
			$QBER_Z$ & 2.5\% &  3.4\%  \\ 
			$QBER_X$ & 2.0\% & 2.2\%  \\ 
			$\phi_{Z}$ & 6.5\% &  8.0\% \\ 
			$\phi_X$ & 5.7\% & 6.3\%  \\ 
			$SKR$ & 4.33kbps & 422.68bps \\
			\hline
		\end{tabular}
\end{table}

Considering a finite-key scenario with a block size of $N = 10^{11}$, the 4D protocol is estimated to deliver secret key rates of 4.33~kbps and 422.68~bps at transmission distances of 200~km and 250~km, respectively. Meanwhile, we theoretically compared the corresponding 2D protocol under the same conditions. It is worth noting that, using the same hardware setup, a single quantum state in the 2D protocol occupies only two time bins, resulting in a repetition rate twice that of the 4D protocol, i.e., 625~MHz. However, as shown in Fig.~\ref{fig4}, the theoretically generated secret key rate of the 2D protocol remains lower across all distances. This advantage of 4D encoding originates from both experimental conditions and the intrinsic properties of high-dimensional states: the doubled information capacity and higher error tolerance reduce error-correction consumption, while the use of high-efficiency, low-dark-count SNSPDs keeps noise counts well below the signal level even at 250~km, preventing QBER saturation in the low-photon-arrival regime. Consequently, even with a doubled repetition rate, the 2D protocol cannot surpass the 4D protocol in overall key generation efficiency. 

In Fig.~\ref{fig4}, we compare our results with other existing typical high-dimensional QKD experiments. In terms of transmission distance, our implementation reaches the furthest distance to date, achieving 250 km. We note that Ref.~\cite{timeengry5} also demonstrated a high-dimensional QKD over 242 km; however, the protocol used there was dispersive optics QKD (DO-QKD), resulting in a relatively lower key rate. In contrast, our results show more than three orders of magnitude higher key rates at similar transmission distances. Ref.~\cite{timebin3} reports a HD time-bin BB84 system with 145 km fiber transmission distance; However, whose key rate is much lower than our system, which on one hand is attributed to high-performance SNSPDs we used, on the other hand is due to their higher system loss at the receiving end. Moreover, Ref.~\cite{timebin3} relies on a free-space unbalanced Mach–Zehnder interferometer for high-dimensional decoding, which suffers from high loss, complex system control and poor stability. On the contrary, our design remains entirely fiber-integrated from state preparation to detection, simplifying packaging and facilitating straightforward extension toward integrated photonic platforms. Ref.~\cite{HDCOW} adopts a high-dimensional coherent one-way (COW) protocol rather than BB84, and shows 40 km transmission distance. It is known that, the COW-type protocol relies on time-of-arrival monitoring and typically require discarding multi-occupied frames for security evaluation, erducing the effective key utilization efficiency, and thus causing lower key rates. In addition, its security analysis remains less mature compared to the well-established BB84 framework. 

For completeness, an overview of recent HD-QKD experiments is summarized, with a detailed comparison presented in Table \ref{tab:HDQKD}, covering aspects such as DoF, distance, protocol, dimension (D), and key rate.

\begin{table*}[tphb]
	\centering
	\caption{Comparison of recent HD-QKD experiments. 
		This work achieves the longest transmission distance (250 km) among all reported HD-QKD implementations.}
	\label{tab:HDQKD}
	\begin{tabularx}{\textwidth}{c l c c c c c c}
		\toprule
		DoF & Related work & Dist. (km) & Channel & Protocol & D & Rep. rate & Key rate \\
		\midrule
		
		\multirow{5}{*}{Time-bin}
		& Islam \emph{et al.} 2017 \cite{timebin1} & 20 & Attenuator & HD-BB84 & 4 & 625M & 26.2 Mbps\\
		& Islam \emph{et al.} 2019 \cite{timebin2} & 20 & Attenuator & HD-BB84 & 8 & 312.5M & 8.65 Mbps\\
		& Vagniluca \emph{et al.} 2020 \cite{timebin3} & 145 & Fiber & HD-BB84 & 4 & 297.5M & 0.42 kbps\\
		& Sulimany \emph{et al.} 2025 \cite{HDCOW} & 40 & Fiber & HD-COW & 8 & 62.5M & 0.14 Mbps\\
		& {\bf This work} & {\bf 250} & {\bf Fiber} & {\bf HD-BB84} & {\bf 4} & {\bf 312.5M} & {\bf 422.68 bps}\\
		\midrule
		
		\multirow{6}{*}{Time-energy}
		& Zhong \emph{et al.} 2015 \cite{timeengry1} & 20 & Fiber & HD-BBM92 & 1024 & — & 2.7 Mbps\\
		& Lee \emph{et al.} 2019 \cite{timeengry2} & 43 & Field fiber & DO-QKD & 4 & 695M & 1.2 Mbps\\
		& Liu \emph{et al.} 2019 \cite{timeengry3} & 20 & Fiber & DO-QKD & 4 & — & 151 kbps\\
		& Chang \emph{et al.} 2024 \cite{timeengry4} & 50 & Fiber & HD-BBM92 & — & — & 0.24 Mbps\\
		& Liu \emph{et al.} 2024 \cite{timeengry5} & 242 & Field fiber & DO-QKD & 3 & — & 0.22 bps\\
		& Yu \emph{et al.} 2025 \cite{timeengry6} & 60 & Fiber & HD-BBM92 & 4 & 1G & 37 bps\\
		\midrule
		
		\multirow{5}{*}{OAM}
		& Mirhosseini \emph{et al.} 2015 \cite{OAM1} & 0.002 & Free space & HD-QKD & 7 & 4K & 6.5 bps\\
		& Sit \emph{et al.} 2017 \cite{OAM11} & 0.3 & Free space & HD-QKD & 4 & — & 0.65 bits/photon\\
		& Larocque \emph{et al.} 2017 \cite{OAM12} & — & Free space & HD-QKD & 10 & — & 0.49 bits/photon\\
		& Bouchard \emph{et al.} 2018 \cite{OAM13} & 0.003 & Underwater & HD-BB84 & 3 & — & 0.307 bits/photon\\
		& Cozzolino \emph{et al.} 2019 \cite{OAM2} & 1.2 & Aircore fiber & HD-BB84 & 4 & 600M & 37.85 kbps\\
		\midrule
		
		\multirow{5}{*}{Path}
		& Ca\~nas \emph{et al.} 2017 \cite{path1} & 0.3 & Multicore fiber & HD-BB84 & 4 & 1K & $4.31\times10^{-6}$ bits/pulse\\
		& Ding \emph{et al.} 2017 \cite{path2} & 0.003 & Multicore fiber & HD-QKD & 4 & 5K & —\\
		& Da \emph{et al.} 2020 \cite{path3} & 2 & Multicore fiber & HD-QKD & 4 & 600M & 29.8 kbps\\
		& Hu \emph{et al.} 2020 \cite{path4} & 11 & Multicore fiber & HD-BB84 & 4 & — & 1.268 bits/coin.\\
		& Hu \emph{et al.} 2021 \cite{path5} & — & Free space & HD-QKD & 8 & — & 3.23 kbps\\
		\midrule
		
		OAM-POL & Wang \emph{et al.} 2019 \cite{OAMpol} & — & Free space & HD-BB84 & 4 & — & 1.849 bpss\\
		OAM-SAM & Wang \emph{et al.} 2021 \cite{OAMSAM} & 25 & Fiber & HD-QKD & 4 & — & 0.201 bpss\\
		Time-bin-Path & Zahidy \emph{et al.} 2024 \cite{timepath} & 52 & Field fiber & HD-BB84 & 4 & 487M & 51.5 kbps\\
		Time-bin-POL & Zhong \emph{et al.} 2024 \cite{timebinpol} & 50 & Fiber & HD-QKD & 4 & 50M & 1.35 bpss\\
		\bottomrule
	\end{tabularx}
	
	\vspace{2mm}
	{\footnotesize \textbf{Notes:} “—” indicates data not reported in the literature.}
\end{table*}

\section{CONCLUSION AND DISCUSSION}
In conclusion, we have presented a directly modulated laser platform for practical high-dimensional BB84 QKD, where injection-locked lasers generate quantum states without external intensity or phase modulators. This design substantially reduces transmitter complexity and insertion loss, enabling stable time-bin and phase encoding with clear advantages for chip-scale integration. Using this platform, we experimentally achieve 4D QKD over 250~km of fiber—the longest transmission distance reported for high-dimensional time-bin schemes—with secret key rates of 4.33~kbps at 200~km and 422.68~bps at 250~km. Under identical hardware conditions, the 4D protocol consistently outperforms 2D operation, confirming the benefits of increased information capacity and improved noise tolerance. These results demonstrate that high-dimensional QKD can be implemented in a compact architecture using commercially accessible components, outlining a feasible route toward scalable long-distance quantum communication networks.

In summary, although time-bin encoding offers significant stability advantages in long-distance fiber-optic communication and the preparation of the corresponding quantum states is relatively straightforward, achieving high-efficiency measurement of time-bin quantum states in experiments remains a substantial challenge. As the dimensionality of time-bin encoding increases, increasingly complex unbalanced interferometer structures are generally required, which can lead to phase instability, timing errors, and reduced system efficiency. Moreover, more post-selection operations are needed in the experiment, resulting in additional intrinsic losses. Increasing the number of time bins per quantum state typically requires reducing the system repetition rate, which decreases the fraction of effectively transmitted quantum signals and further limits overall performance.

Looking ahead, overcoming these challenges will rely on the development of more efficient integrated photonic platforms, particularly those capable of generating brighter entangled sources within on-chip waveguides. With continuous advances in integrated photonics and high-performance single-photon detectors—especially with improved picosecond-level resolution for time-bin interval discrimination—time-bin–encoded HD-QKD is expected to overcome its current performance bottlenecks. Furthermore, given the maturity of time-bin modulation technology in both classical and quantum communication, hybrid encoding schemes that combine time-bin and OAM degrees of freedom merit further investigation. Such joint encoding has the potential to significantly enhance system dimensionality and information capacity. Nevertheless, since time-bin encoding inherently consumes temporal resources to increase dimensionality, trade-offs between system bandwidth and overall performance must be carefully considered. Future progress will also depend on achieving precise phase control and low-loss designs that enable an increased number of time bins without compromising the system repetition rate, thereby paving the way for practical deployment of HD-QKD in large-scale optical fiber networks with improved transmission rates and enhanced security.

\section*{Funding} This work is supported by the Natural Science Foundation of Jiangsu Province (No. BE2022071), and the National Natural Science Foundation of China (NSFC) (Nos. 12074194, 62101285, and 62471248).

\section*{Disclosures}  The authors declare no conflicts of interest.

\section*{Data availability} Data underlying the results presented in this paper can be obtained from the authors upon a reasonable request.
\bibliography{HD.bib}
\end{document}